\documentclass[12pt]{iopart}

\usepackage{graphicx}

\begin{document}

\title[Measurements of the superconducting fluctuations in optimally-doped BaFe$_{2-x}$Ni$_x$As$_2$]
{Measurements of the superconducting fluctuations in optimally-doped BaFe$_{2-x}$Ni$_x$As$_2$ under high magnetic fields: Probing the 3D anisotropic Ginzburg-Landau approach}

\author{R I Rey$^1$, A Ramos-\'Alvarez$^1$, C Carballeira$^1$, J Mosqueira$^1$, F Vidal$^1$, S~Salem-Sugui~Jr.$^2$, A D Alvarenga$^3$, Rui Zhang$^4$, Huiqian Luo$^4$}

\address{$^1$LBTS, Facultade de F\'isica, Universidade de Santiago de Compostela, E-15782 Santiago de Compostela, Spain}

\address{$^2$Instituto de Fisica, Universidade Federal do Rio de Janeiro, 21941-972 Rio de Janeiro, RJ, Brazil}

\address{$^3$Instituto Nacional de Metrologia Qualidade e Tecnologia, 25250-020 Duque de Caxias, RJ, Brazil}

\address{$^4$Beijing National Laboratory for Condensed Matter Physics, Institute of Physics, Chinese Academy of Sciences, Beijing 100190, China}

\ead{j.mosqueira@usc.es}

\begin{abstract}
The superconducting fluctuations well inside the normal state of Fe-based superconductors were experimentally studied through the in-plane paraconductivity in several high-quality optimally-doped BaFe$_{2-x}$Ni$_x$As$_2$ crystals. These measurements were performed in magnetic fields with amplitudes up to 14 T, and different orientations relative to the crystals $c$ axis ($\theta=0^\circ$, 53$^\circ$, and 90$^\circ$). The results allowed a stringent check of the applicability of a recently proposed Ginzburg-Landau approach for the fluctuation electrical conductivity of 3D anisotropic materials in presence of finite applied magnetic fields.
\end{abstract}

\pacs{74.25.Ha, 74.40.-n, 74.70.Xa}
\submitto{\SUST}
\maketitle

\section{Introduction}

The high critical temperatures ($T_c$) of Fe-based superconductors (FeSC), and the unconventional mechanism for their superconductivity (with a pairing probably mediated by spin fluctuations and involving several bands) have generated an enormous interest for these materials in the last few years.\cite{reviews} A central aspect of their phenomenology is the effect of superconducting fluctuations around $T_c$.\cite{tinkham} Mainly due to the short coherence length and high-$T_c$ values of these materials,\cite{reviews} these effects are enhanced with respect to conventional low-$T_c$ superconductors. In fact, the Ginzburg number characterizing the width of the critical fluctuation region around $T_c$, is in FeSC halfway the ones found in conventional low-$T_c$ superconductors and high-$T_c$ cuprates (HTSC).\cite{pallecchi}

\begin{figure}[t]
\begin{center}
\includegraphics[scale=.6]{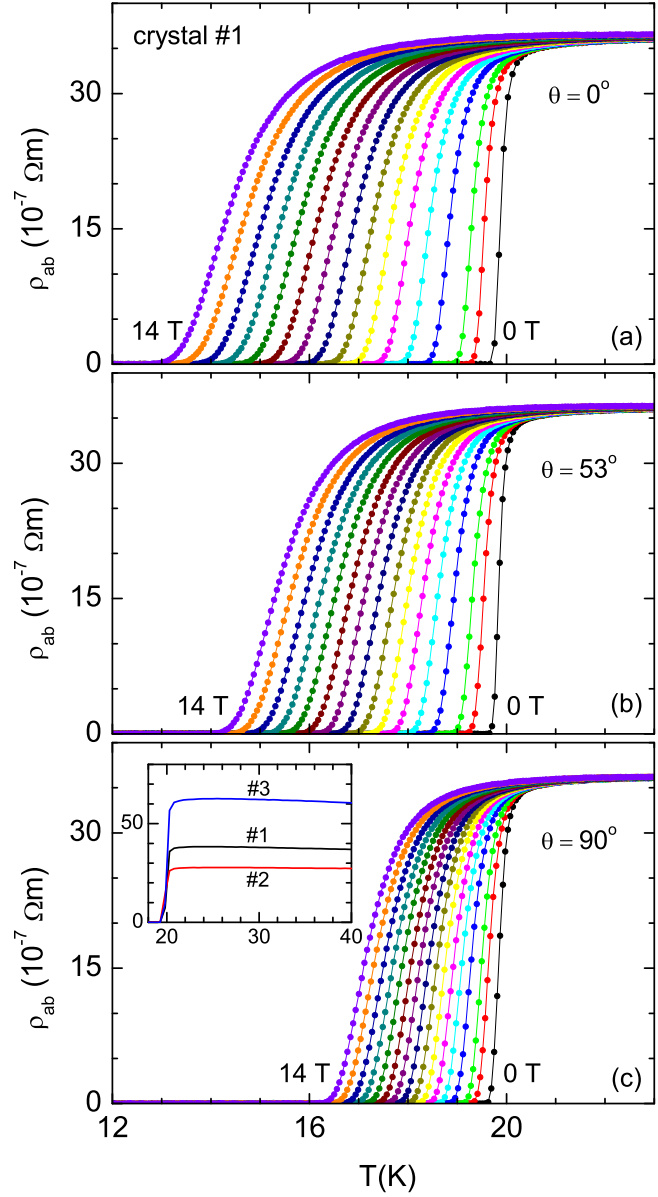}
\caption{Temperature dependence of the resistivity around $T_c$ for crystal \#1. These measurements were performed in presence of different magnetic field amplitudes (0, 0.5~T, and from 1 to 14~T in steps of 1~T) and orientations ($\theta=0^\circ,53^\circ$, and $90^\circ$) relative to the crystal $c$-axis. Inset in (c): overview up to $\sim2T_c$ of the resistivity in absence of field for all the crystals studied.}
\label{rhoTH}
\end{center}
\end{figure}

In addition to their intrinsic interest, the superconducting fluctuation effects are a very useful tool to characterize the nature of a superconducting transition and to obtain material parameters,\cite{tinkham} so that different works have already addressed their study in FeSC through observables like the magnetization, specific heat or the electric conductivity.\cite{pallecchi,fanfarillo,salemsugui,choi,putti,liuPLA10,tesanovic,pandyaSST10,kim,liuSSC11,pandyaSST11,mosqueira,welp,liu2,prando,song,marra,rullier,rey,critical,mosqueiraJSNM13} 
However, some fundamental aspects of the phenomenology of the fluctuation effects in these materials are still debated. One of them is their dimensionality. In these materials the transverse coherence length amplitude $\xi_c(0)$ is close to the Fe layers periodicity length, $s$. Thus, depending on the particular compound studied, some works report a two-dimensional (2D) behavior\cite{pallecchi,pandyaSST10,rullier} similar to the one found in highly anisotropic HTSC,\cite{reviewHTSC} while others find three-dimensional (3D) characteristics,\cite{fanfarillo,salemsugui,choi,putti,kim,liuSSC11,mosqueira,welp,liu2,marra,rey,critical,mosqueiraJSNM13} or even a 3D-2D transition\cite{liuPLA10,pandyaSST11,song} when
increasing the temperature above $T_{c}$ 
(as in optimally-doped YBa$_2$Cu$_3$O$_{7-\delta}$)\cite{reviewHTSC}.\footnote{A theory for the effect of critical fluctuations around the $T_c(H)$ line on different observables in superconductors with intermediate 2D-3D characteristics, was developed in Ref.~\cite{tesanovic}}
But it was recently reported that the fluctuation electrical conductivity above $T_{c}$ of clean LiFeAs crystals seemingly follows a well defined 2D behavior (in both the amplitude and the reduced-temperature dependence), despite that for this compound $\xi_c(0)\approx1.6$~nm is much larger than the Fe-layers periodicity length ($s=0.636$~nm).\cite{rullier} 
This surprising result led the authors of Ref.~\cite{rullier} to propose that in these multiband superconductors the fluctuating pairs above $T_c$ may be driven by a single 2D band. 
Other interesting issues that deserve attention are the possible presence of \textit{phase fluctuations} (which effect was possibly observed near $T_c$ at low field amplitudes in the SmFeAsO$_{0.8}$F$_{0.2}$ \cite{prando}, but also in members of the less anisotropic 122 family as Ba$_{1-x}$K$_x$Fe$_2$As$_2$ \cite{salemsugui} and Ba(Fe$_{1-x}$Rh$_x$)$_2$As$_2$ \cite{bossoni}), or the behavior of fluctuation effects in the short wavelength regime appearing at high reduced magnetic fields or temperatures.

As a contribution for the understanding of the above mentioned issues, here we present detailed measurements of the fluctuation-induced in-plane electric conductivity ($\Delta\sigma_{ab}$) in several high quality BaFe$_{2-x}$Ni$_x$As$_2$ crystals with doping levels near the optimal one ($x\approx0.1$). These experiments were performed in magnetic fields ($H$) up to 14~T applied with different angles $\theta$ relative to the crystals $c$-axis ($\theta=0$, 53, and 90 degrees), thus extending previous measurements in the same compound with $H\perp ab$ up to 9~T.\cite{rey,critical} 
The large fields used here allow to deeply penetrate into the so-called Prange fluctuation regime, and to perform a stringent check of the applicability of a recently proposed generalization of the classic Aslamazov-Larkin (AL) results to finite fields through a 3D-anisotropic Ginzburg-Landau approach.\cite{rey} In turn, the use of different magnetic field orientations provides an important consistency test of the analysis and allows to obtain precise information about the system dimensionality, basic superconducting parameters (as the coherence lengths and the anisotropy factor), and the angular dependence of the upper critical field, at present another debated issue in these materials.\cite{angularHc2}

\section{Experimental details and results}

We studied three BaFe$_{2-x}$Ni$_x$As$_2$ single crystals with nominal doping levels near the optimum one, two with $x=0.096$ (\#1 and \#2) and one with $x=0.098$ (\#3). Their sizes are typically $1.5\times1.0\times0.3$ mm$^3$, being the $c$-axis of the tetragonal structure ($a=b=3.96$~\r{A}, $c=12.77$~\r{A}) perpendicular to their largest face. Details of their growth procedure and a thorough characterization may be found in Ref.~\cite{growth}.

\begin{table}[b]
\begin{center}
\begin{tabular}{ccccccc}
\hline
Crystal  & $T_c$(K) & $\Delta T_c$(K) & $\xi_c$ (nm) & $\xi_{ab}$ (nm) & $\gamma$ \\ 
        &   &  & $\pm$6\% & $\pm$2\% & $\pm$8\%\\
\hline  
\#1  & 19.8 & 0.3 & 1.28 & 2.57 & 2.00  \\  
\#2 & 19.7 & 0.2 & 1.28 & 2.54 & 1.98 \\ 
\#3 & 19.8 & 0.2 & 1.24 & 2.52 & 2.02 \\  
\hline
\end{tabular}
\end{center}
\caption{Summary of the superconducting parameters resulting from the analysis.}
\end{table}

The resistivity along the $ab$ layers, $\rho_{ab}$, was measured with a Quantum Design's Physical Property Measurement System (PPMS) in presence of magnetic fields up to 14~T with different orientations relative to the $c$-axis ($\theta=0$, 53, and 90 degrees). For that we used a standard four-probe method with a low contact resistance (less than 1~Ohm) and a current of 1 mA. The data were obtained by sweeping the temperature at a rate of 0.3 K/min. An example of the $\rho_{ab}(T)_{H,\theta}$ behavior (corresponding to crystal \#1) around $T_c$ is presented in Fig.~1. The $T_c$ value was determined from the transition midpoint for the $H=0$ data, and the transition width was estimated as $\Delta T_c\approx T_c-T(\rho=0)$. The corresponding values for the three samples studied are compiled in Table~1. The small $\Delta T_c/T_c$ values (about $10^{-2}$) confirm the excellent stoichiometric quality of the crystals.
An overview of $\rho_{ab}(T)$ in absence of field and up to $\sim 2T_c$ is presented in the inset of Fig.~1(c) for all samples studied. As it may be seen, $\rho_{ab}$ is almost temperature independent from few degrees above $T_c$ up to $2T_c$. This is an important experimental advantage to determine the conductivity induced by superconducting fluctuations (or \textit{paraconductivity}), which is given by 
\begin{equation}
\Delta\sigma_{ab}(T,H)=\frac{1}{\rho_{ab}(T,H)}-\frac{1}{\rho_{ab,B}(T,H)},
\end{equation} 
where $\rho_{ab,B}$ is the normal-state or \textit{background} contribution. The procedure to estimate $\rho_{ab,B}$ is illustrated in the example of Fig.~2, also corresponding to crystal \#1: In the region $26-30$~K (corresponding to $1.3-1.5$~$T_c$, where fluctuation effects are expected to be negligible \cite{rey}) the resistivity is linear with the temperature up to the largest field used in the experiments. Besides, as it is shown in the inset of that figure, the magnetoresistivity in the normal state is roughly quadratic in the applied magnetic field. This allows to parametrize the background resistivity as
\begin{equation}
\rho_{ab,B}(T,H)=\alpha(H)+\beta(H)T
\end{equation}
where
\begin{eqnarray}
&&\alpha(H)=a_1+a_2H^2\nonumber\\
&&\beta(H)=b_1+b_2H^2
\end{eqnarray}
The coefficients $a_1$, $a_2$, $b_1$, and $b_2$ were obtained by linear fittings to the $\rho(T)$ curves measured with $\mu_0H=0$ and 14~T. An example (corresponding to crystal \#1) of the resulting $\Delta\sigma_{ab}$ dependence on the reduced temperature, $\varepsilon\equiv\ln(T/T_c)$, is presented in Fig.~3.

\begin{figure}[t]
\begin{center}
\includegraphics[scale=.6]{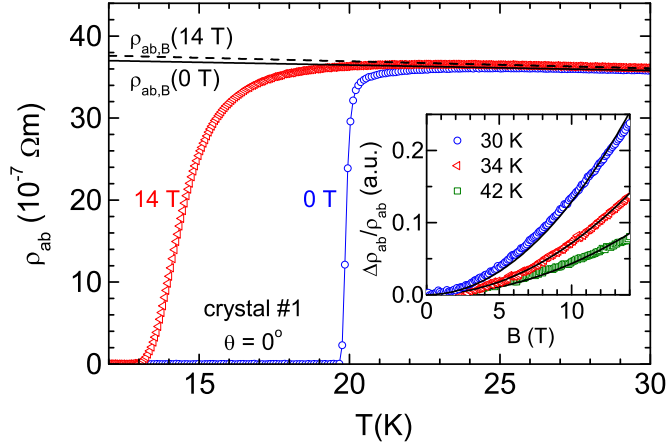}
\caption{Temperature dependence of the resistivity around $T_c$ for crystal \#1, measured with $\mu_0H=0$ and 14~T perpendicular to the $ab$ layers ($\theta=0^\circ$). The corresponding normal-state (or background) contributions (lines) were obtained by a linear fit above 26~K (i.e., 1.3~$T_c$), where fluctuation effects are expected to be negligible. Inset: field dependence of the magnetoresistivity, $[\rho_{ab}(B)-\rho_{ab}(0)]/\rho_{ab}(0)$, for several temperatures above $T_c$. The lines are fits to a quadratic form $\Delta\rho_{ab}/\rho_{ab}={\rm cte.}B^2$.}
\label{backs}
\end{center}
\end{figure}

\section{Data analysis}

\subsection{In-plane paraconductivity in the low-field limit}

\begin{figure}[t]
\begin{center}
\includegraphics[scale=.5]{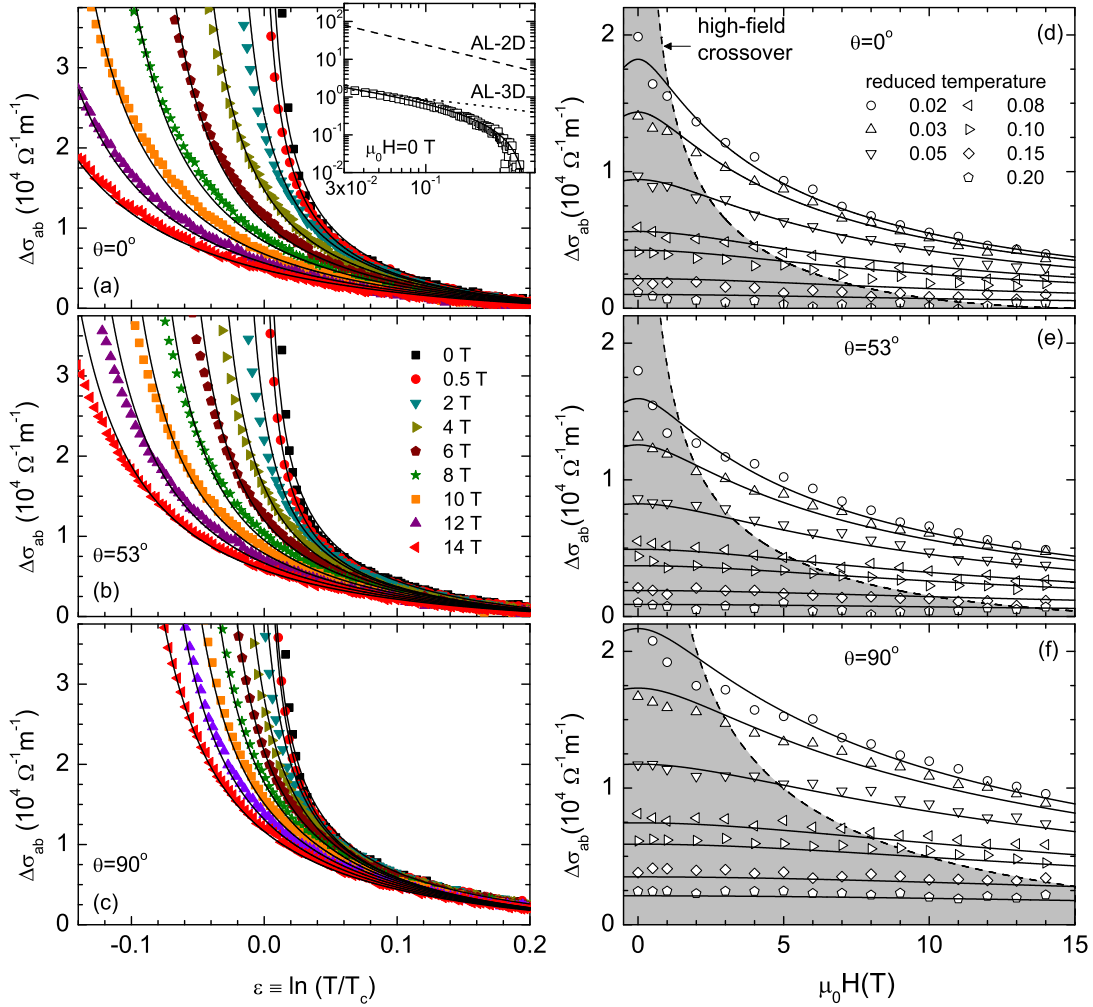}
\caption{Example for crystal \#1 of the $\Delta\sigma_{ab}$ dependence on the reduced temperature (a-c) and on the magnetic field amplitude (d-f). The indicated $\theta$ values represent the angle between the applied magnetic field and the crystal $c$ axis. The lines are the best fits of Eq.~(\ref{prange}) using only three free parameters for the entire set of data of each field orientation: $\xi_c$, $H_{c2}(\theta)$, and $C$. The dashed lines in (d-f) represent the crossover to the Prange regime, according to the criterium $h=\varepsilon$. Inset in (a): $\log$-$\log$ plot of the $\varepsilon$-dependence of $\Delta\sigma_{ab}$ in absence of an applied field. Solid and dotted lines are the best fits of Eq.~(\ref{prange}) and, respectively, the 3D-AL approach, Eq.~(\ref{AL}) (this last for $\varepsilon<0.1$, where short-wavelength effects are expected to be negligible). The dashed line is the prediction of the 2D-AL approach, Eq.~(\ref{2DAL}).}
\label{Ds}
\end{center}
\end{figure}

In the absence of a magnetic field and for temperatures close to $T_c$, it is expected that $\Delta\sigma_{ab}$ will follow the classical Aslamazov-Larkin result, which for 3D superconductors may be written as\cite{AL}
\begin{equation}
\Delta\sigma_{ab}=\frac{e^2}{32\hbar\xi_c}\varepsilon^{-1/2},
\label{AL}
\end{equation} 
where $e$ is the electron charge, $\hbar$ is the reduced Planck constant, and $\xi_c$ is the $c$-axis coherence length amplitude. As it may be seen in the inset of Fig.~\ref{Ds}, for reduced-temperatures below $\varepsilon\approx0.1$ a critical exponent close to $-1/2$ is observed, in agreement with Eq.~(\ref{AL}). Above this $\varepsilon$-value a rapid falloff of the fluctuation effects is observed and a well-defined critical exponent is no longer observed, a behavior that may be attributed to short-wavelength fluctuation effects.\cite{tinkham,EPLcutoff,IJMPVidal} For completeness, in the same inset we present the prediction of the 2D-AL result
\begin{equation}
\Delta\sigma_{ab}=\frac{e^2}{16\hbar s}\varepsilon^{-1},
\label{2DAL}
\end{equation} 
where $s=6.38$~\r{A} is the Fe-As layers periodicity length. As it may be clearly seen, it overestimates the experimental data by almost two orders of magnitude, which is well beyond the experimental uncertainties, including those associated to the determination of the normal-state background. 

In the presence of a finite magnetic field, roughly above the so-called \textit{ghost critical field} $H^*(T)$ (which is the symmetric above $T_c$ of the corresponding $H_{c2}(T)$ line \cite{ghost}), $\Delta\sigma_{ab}$ is expected to be significantly reduced with respect to Eq.~(\ref{AL}).\cite{tinkham,PRL00,breakdown} As it may be seen in Fig.~\ref{Ds}, in particular in the $\Delta\sigma_{ab}(H)_\varepsilon$ representation of panels (d-f), such a reduction is clearly observed with the field amplitudes used in our experiments and, as expected, it is more prominent for temperatures close to $T_c$ (i.e., for $\varepsilon\to0$). It is also noticeable the dependence on the field orientation relative to the crystals $c$-axis, which is a direct consequence of the anisotropy of the upper critical field in the studied compound (see below).

\subsection{Comparison with the GL approach for the finite-field or Prange regime}

In what follows we will present an analysis of the experimental data in terms of the 3D-anisotropic Ginzburg-Landau (GL) approach developed in Ref.~\cite{rey}. It adapts to this dimensional case the model proposed by A. Schmid, which is based on a combination of the standard Gaussian GL-expression of the thermally-averaged current density with the generalized Langevin equation of the order parameter.\cite{schmidt} Since the energy of the fluctuation modes increases with $H$, this finite-field approach includes an energy cutoff in the fluctuations spectrum as proposed in Refs.~\cite{EPLcutoff,IJMPVidal}. For $H$ perpendicular to the $ab$ layers, it leads to
\begin{equation}
\Delta\sigma_{ab}=\frac{e^2}{32\hbar \pi\xi_c}\sqrt{\frac{2}{h}}\int_0^{\sqrt{\frac{C-\varepsilon}{2h}}}{\rm d}x
\left[\psi^1\left(\frac{\varepsilon+h}{2h}+x^2\right)-\psi^1\left(\frac{C+h}{2h}+x^2\right)
\right],
\label{prange}
\end{equation}
where $h=H/H_{c2}^\perp$, $H_{c2}^\perp$ is the linear extrapolation to $T=0$~K of the upper critical field for $H\perp ab$ (i.e., $\theta=0^\circ$), and $C$ is a cutoff constant which value is expected to be about $\sim0.5$.\cite{EPLcutoff,IJMPVidal} In the zero-field limit (for $h\ll \varepsilon,C$), Eq.~(\ref{prange}) is transformed into 
\begin{equation}
\Delta\sigma_{ab}=\frac{e^2}{16\hbar \pi\xi_c}\left(\frac{{\rm arctan}\sqrt{\frac{C-\varepsilon}{\varepsilon}}}{\sqrt{\varepsilon}}-\frac{{\rm arctan}\sqrt{\frac{C-\varepsilon}{C}}}{\sqrt{C}}\right),
\label{schmidt}
\end{equation}
which at low reduced temperatures ($\varepsilon\ll C$) reduces to the conventional AL expression, Eq.~(\ref{AL}).
Following the scaling transformation for anisotropic materials developed in Refs.~\cite{Klemm80,Blatter92,Hao92}, Eq.~(\ref{prange}) may be generalized to an arbitrary field orientation by just replacing $h$ by 
\begin{equation}
h_\theta=\frac{H}{H_{c2}(\theta)},
\label{reducedh}
\end{equation}
being $H_{c2}(\theta)$ the upper critical field (linearly extrapolated to $T=0$~K) for an arbitrary angle $\theta$ between ${\bf H}$ and the crystal $c$ axis.  

\begin{figure}[t]
\begin{center}
\includegraphics[scale=.6]{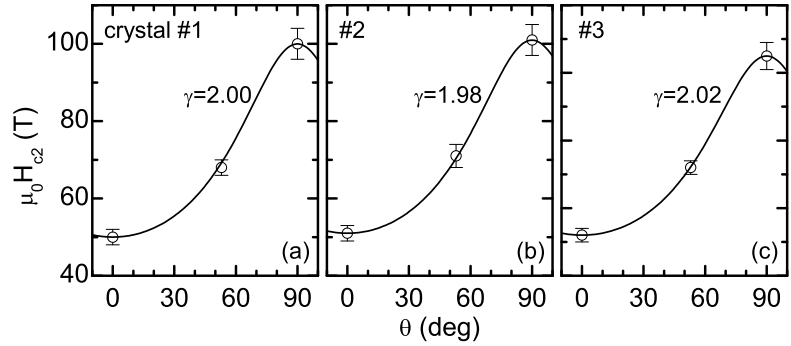}
\caption{Dependence of $H_{c2}$ on the angle between the applied field and the crystals $c$ axis. These data result from the analysis of $\Delta\sigma_{ab}(T,H)$ in the normal state in terms of Eq.~(\ref{prange}). The lines correspond to the 3D-anisotropic GL expression, Eq.~(\ref{Hc2angular}), evaluated by using the parameters in Table~1.}
\label{plotangular}
\end{center}
\end{figure}

\begin{figure}[t]
\begin{center}
\includegraphics[scale=.6]{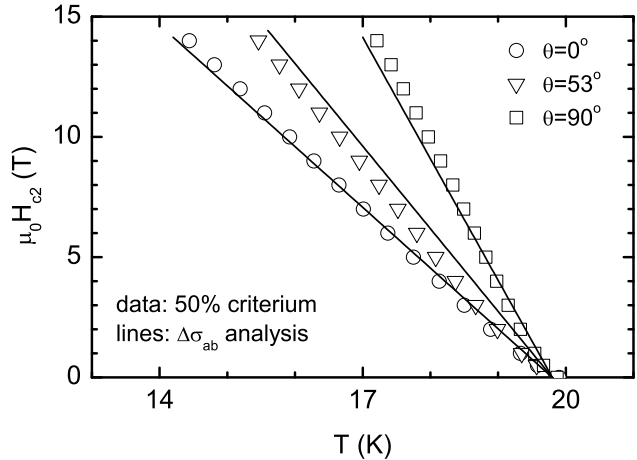}
\caption{Temperature dependence of the upper critical field for the three field orientations studied, as results from the analysis of fluctuation effects (lines) and from the 50\% criterium (data points). The data of this example correspond to crystal \#1. See the main text for details.}
\label{criterios}
\end{center}
\end{figure}

Equation (\ref{prange}) is fitted to the complete set of data for each field orientation with only three free parameters: the upper critical field $H_{c2}(\theta)$, the amplitude term (directly related to $\xi_c$), and the cutoff constant $C$. As it may be seen in Figs.~\ref{Ds}(a-c) the agreement is excellent, extending down to a field-dependent temperature below $T_c$ that may be close to the upper bound of the \textit{critical region}.\footnote{In this region fluctuation effects are so important that the Gaussian approximation [used to derive Eq.~(\ref{prange})] is no longer applicable, see Ref.~\cite{tinkham}.} The agreement is also excellent in the $\Delta\sigma_{ab}(H)$ representation of Figs.~\ref{Ds}(d-f), which is focused on temperatures above $T_c$. 
The resulting $H_{c2}(\theta)$ values are presented in Fig.~\ref{plotangular} for all samples studied. These data follow the behavior expected for 3D-anisotropic materials (solid lines) \cite{tinkhamHTSC},
\begin{equation}
H_{c2}(\theta)=\left[\frac{\cos^2\theta}{H_{c2}^2(0^\circ)}+\frac{\sin^2\theta}{H_{c2}^2(90^\circ)}\right]^{-1/2},
\label{Hc2angular}
\end{equation}
which represents an important consistency check of the present results. The $H_{c2}(0^\circ)$ and $H_{c2}(90^\circ)$ values are within the ones obtained in the literature in the same material from the shift of the resistive transition induced by the field,\cite{sun,tao,ni,shahbazi} although the rounding associated to fluctuation effects makes this procedure strongly dependent on the criterion used (generally a given \% of the normal-state resistivity). In the present case, as it may be seen in Fig.~\ref{criterios}, the 50\% criterion gives $H_{c2}$ values in good agreement with the ones resulting from the analysis of fluctuation effects. 

As the $\Delta\sigma_{ab}$ amplitude may be affected by the uncertainties associated with the finite size of the electrical contacts and with the crystals geometry, the amplitude term in Eq.~(\ref{prange}) is not used to determine $\xi_c$. Instead, the GL coherence length amplitudes are obtained from the $H_{c2}(\theta)$ values in Fig.~\ref{plotangular} according to
\begin{equation}
\xi_{ab}=\left[\frac{\phi_0}{2\pi\mu_0H_{c2}(0^\circ)}\right]^{1/2},
\end{equation}and
\begin{equation}
\xi_c=\xi_{ab}/\gamma,
\end{equation}
where the anisotropy factor $\gamma$ is obtained from the ratio
\begin{equation}
\gamma=\frac{H_{c2}(90^\circ)}{H_{c2}(0^\circ)}.
\end{equation}
The values corresponding to each sample are compiled in Table~1. 

Regarding the cutoff constant, it resulted to be in the range $C=0.35\pm0.05$ for all samples. This value is close to the ones found in previous works about fluctuation effects above $T_c$ in FeSC, in particular on the paraconductivity at high-$\varepsilon$ values in the same compound,\cite{rey} and on the precursor diamagnetism in optimally-doped Ba$_{1-x}$K$_x$Fe$_2$As$_2$.\cite{mosqueira} It is also close to the cutoff constant found in other superconducting families, including high-$T_c$ cuprates,\cite{intrinsic,Tl2223,TlPb1212,YBCO} low-$T_c$ metallic elements and alloys,\cite{PbNb,PbIn} and compounds like MgB$_2$ or NbSe$_2$.\cite{MgB2,NbSe2} Our results confirm the proposal in Refs.~\cite{EPLcutoff,IJMPVidal} about an universal $C$ value close to $\sim0.5$, associated with the limits encountered at high-$\varepsilon$ or $h$ to the shrinkage of the superconducting wavefunction to lengths of the order of the pairs size.

Let us finally comment on the applicability of a GL approach to a two-band superconductor as BaFe$_{2-x}$Ni$_x$As$_2$. In principle, the analysis of fluctuation effects in multiband superconductors, in particular when they involve two or more weakly-coupled bands with different anisotropy, would require a specific multiband functional that takes into account the non-local effects arising from having an effective coherence length in one of the crystallographic directions of the system much smaller than the associated to one of the bands. It has been proposed that this is the case of MgB$_2$ \cite{varlamov}, although a good description of fluctuation effects in terms of GL approaches was also found for this compound \cite{MgB2}. The applicability of a GL approach to BaFe$_{2-x}$Ni$_x$As$_2$ would suggest that the interband coupling in this compound is larger than that in MgB$_2$. This is consistent with the fact that the relative band interaction constant defined in Ref.~\cite{varlamov}, $S_{12}$, is in MgB$_2$ of the order of $0.035$, while in optimally doped BaFe$_{2-x}$Ni$_x$As$_2$ we find $S_{12}\simeq0.134$ (i.e., four times larger) by using the coupling parameters reported in Ref.~\cite{lambda}.

\begin{figure}[b]
\begin{center}
\includegraphics[scale=.6]{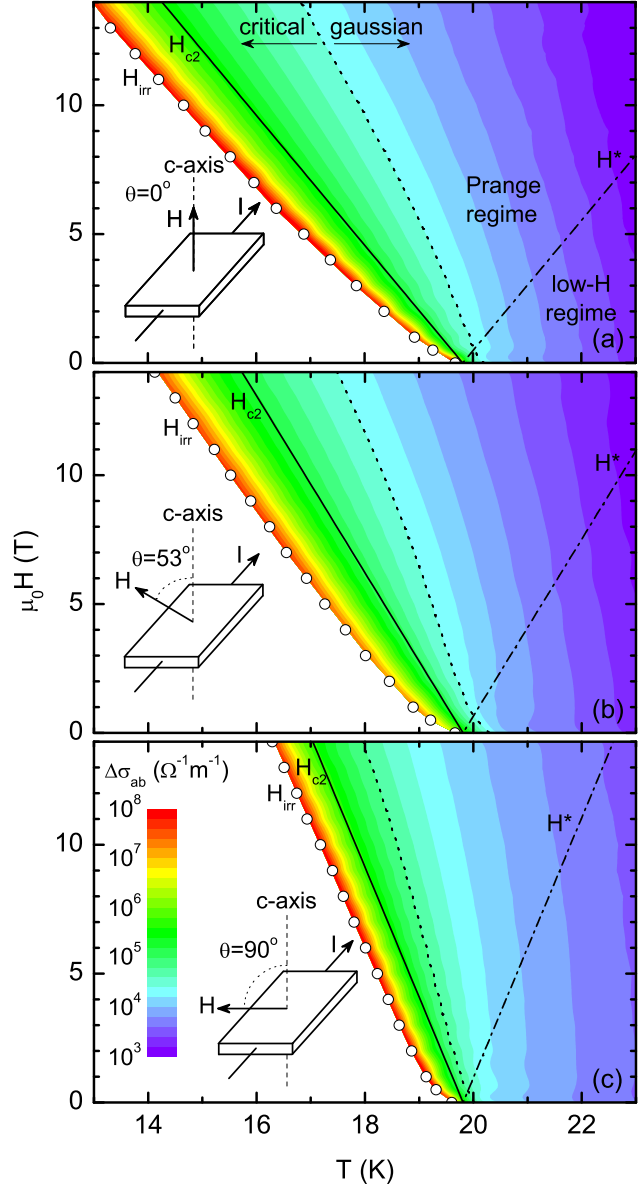}
\caption{Experimental $H-T$ phase diagram for crystal \#1, showing the $\Delta\sigma_{ab}$ amplitude for the three {\bf H} orientations studied. The circles indicate where the resistivity vanishes. The dotted line is the observed limit of applicability of Eq.~(\ref{prange}), and roughly separates the \textit{gaussian} and \textit{critical} fluctuation regimes. The solid line is the upper critical field resulting from the analysis of $\Delta\sigma_{ab}(T,H)$ in terms of Eq.~(\ref{prange}). The dot-dashed line is the so-called \textit{ghost field} (the symmetric above $T_c$ of the upper critical field), above which finite field effects are expected to be relevant.}
\label{phase}
\end{center}
\end{figure}

\begin{figure}[b]
\begin{center}
\includegraphics[scale=.6]{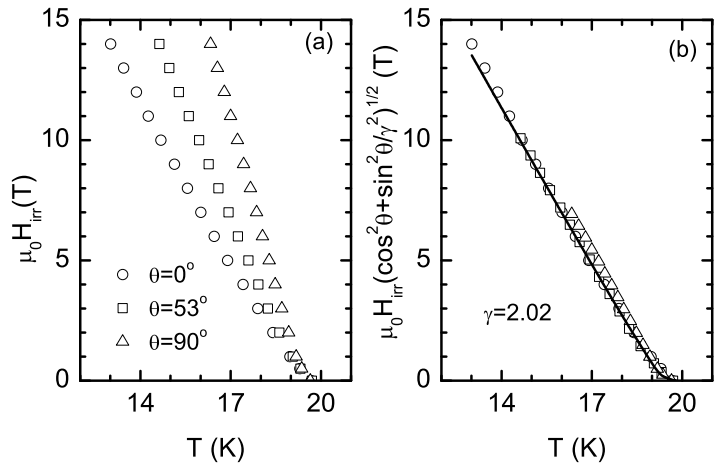}
\caption{(a) Temperature dependence of the irreversibility field $H_{\rm irr}(T)$ for the three orientations of the applied field (this example corresponds to crystal \#3). These data were obtained from the temperatures at which $\rho_{ab}(T)_{H,\theta}$ vanishes. (b) 3D-anisotropic GL scaling of the $H_{\rm irr}(T)_{H,\theta}$ data, evaluated by using the $\gamma$ value in Table~1. The line is the best fit of the theoretical approach presented in Ref.~\cite{baruch} (see main text for details).}
\label{irrline}
\end{center}
\end{figure}

\subsection{$H-T$ phase diagram for $\Delta\sigma_{ab}$}

The large number of measured $\Delta\sigma_{ab}$ isofields allowed to plot detailed $H-T$ phase diagrams of the $\Delta\sigma_{ab}$ amplitude for the three field orientations studied. An example for crystal \#1 is shown in Fig.~\ref{phase}. The solid line is the upper critical field, as obtained from the $T_c$ and $H_{c2}$ values in Fig.~\ref{plotangular} by assuming a linear temperature dependence close to the transition. The dotted line represents the experimental limit of applicability of Eq.~(\ref{prange}) which, as commented above, may be close to the onset of the \textit{critical region}. In these phase diagrams finite field effects may be seen as deviations from the verticality of the iso-$\Delta\sigma_{ab}$ curves. These effects are more prominent for $H\parallel c$ and, as expected, appear for fields roughly above the corresponding \textit{ghost field}, $H^*(T)$. Finally, the circles indicate the points at which $\rho_{ab}(T)_H$ falls below the noise level, which are expected to be close to the irreversibility line, $H_{\rm irr}(T)$. It is worth noting that the dependence of  $H_{\rm irr}(T)$ on the orientation of the applied magnetic field provides a further check of the applicability of the 3D anisotropic GL approach to the compound under study. In fact, by approximating the irreversibility line by the melting line, according to Ref.~\cite{Hao92} it is expected that
\begin{equation}
T_{\rm irr}({\bf H})=T_{\rm irr}\left(\frac{H}{H_{c2}(\theta)}\right).
\end{equation}
Then, taking into account the above $H_{c2}(\theta)$ dependence, the $H_{\rm irr}(T)_\theta$ lines should scale when normalized by $(\cos^2\theta+\gamma^{-2}\sin^2\theta)^{-1/2}$. As it may be seen in Fig.~\ref{irrline}, such a scaling is observed when using the $\gamma$ values in Table~1. 
Just for completeness, note that the irreversibility line for $\theta=0^\circ$ follows the temperature dependence predicted in Ref.~\cite{baruch}, which was obtained within a 3D disordered Ginzburg-Landau model:
\begin{equation}
1-t-h+2\left[\frac{n_p(1-t)^2h}{4\pi}\right]^{2/3}\left[\frac{3}{2}-\frac{4\pi t\sqrt{2Gi}}{n_p(1-t)^2}\right]=0,
\label{baruch}
\end{equation}
where $t=T/T_c$, $h=H/H_{c2}(\theta=0^\circ)$, and $n_p$ and $Gi$ are fitting parameters defined in Ref.~\cite{baruch} representing the disorder and the strength of thermal fluctuations, respectively. Values obtained from the fitting are $\mu_0H_{c2}(0^\circ)=42~T$, $n_p=0.006$ and $Gi=10^{-6}$. The value of $n_p$ suggests that disorder is important in the reversible region below the $H_{c2}(T)$ line.

\subsection{Comparison with recent works}

In a recent work by Rullier-Albenque \textit{et al}. (Ref.~\cite{rullier}) it is reported that the paraconductivity of clean LiFeAs samples is 2D in nature, in spite that the superconducting parameters of this compound (similar to the ones of BaFe$_{2-x}$Ni$_x$As$_2$) would suggest a 3D behavior: $\xi_c\sim1.6$~nm is much larger than the interlayer distance, $s=0.636$~nm. Here we show that this conflicting result may be an artifact associated to the procedure used to determine the normal-state contribution. First of all, let us note that in clean crystals the same fluctuation effects (i.e., the same $\Delta\sigma_{ab}$) lead to 
a much weaker resistivity rounding than in dirty samples (with a much larger background resistivity): the reason is that not too close to $T_c$ the change in the electrical resistivity due to superconducting fluctuations may be approximated by
\begin{equation}
\Delta\rho_{ab}\approx\rho_{ab,B}^2\Delta\sigma_{ab}.
\label{deltarho}
\end{equation}
In the clean crystals used in Ref.~\cite{rullier}, $\rho_{ab,B}\approx4\times10^{-8}\;\Omega$m just above $T_c$. Subsequently, in the case of 3D fluctuations, at intermediate reduced temperatures (e.g., $\varepsilon=0.1$) the relative change in $\rho_{ab}$ is expected to be about 0.06\%. Even in the case of 2D fluctuations, for which $\Delta\sigma_{ab}$ is given by Eq.~(\ref{2DAL}), the relative change in $\rho_{ab}$ is expected to be smaller than 1\%. Indeed, the observation of fluctuation effects in the resistivity of so clean crystals would require an extraordinary precision in the determination of the background, whatever the procedure used. In Ref.~\cite{rullier} $\rho_{ab,B}$ is estimated by allegedly quenching fluctuation effects with magnetic fields typically above 10~T. However, it has been shown that fluctuations above $T_c$ survive up to fields of the order of $T_c(dH_{c2}/dT)_{T_c}$.\cite{breakdown} As for LiFeAs this quantity is about 28~T \cite{cho}, it is likely that fluctuation effects are still present above 10 T. 

Nevertheless, in Ref.~\cite{rullier} it is found a good agreement (without free parameters) between Eq.~(\ref{2DAL}) and the data obtained in one of the samples (named FP2). However, there is a large difference with the results obtained in the other sample (FP1): the $\Delta\sigma_{ab}$ data at 18.8 K ($\varepsilon\approx0.09$), not included in Fig.~4 of Ref.~\cite{rullier} but available from the data in Fig.~2(b), is a factor 2.5 larger than the one for sample FP2. The difference cannot be attributed to a wider superconducting transition [as it is shown in Fig.~1(a) both samples present similar transition widths], and suggests that the agreement of Eq.~(\ref{2DAL}) with sample FP1 could be accidental. If the fluctuation effects in LiFeAs were actually 3D in nature, the arguments given in Ref.~\cite{rullier} supporting a pure $H^2$ behavior for transverse magnetoresistivity in this compound, should be revised.

\section{Conclusions}

We have presented detailed measurements of the conductivity induced by superconducting fluctuations just above the superconducting transition of three high-quality optimally-doped BaFe$_{2-x}$Ni$_x$As$_2$ single crystals. These measurements were performed with magnetic fields up to 14 T, which allowed to deeply penetrate into the finite-field (or Prange) fluctuation regime. The magnetic field was applied with different orientations with respect to the crystals $c$ axis ($\theta=0^\circ$, 53$^\circ$, and 90$^\circ$), allowing to investigate the anisotropy of fluctuation effects. The analysis of the experimental data lead to solid evidence of the applicability to these compounds of a recently published Gaussian GL approach for 3D anisotropic superconductors in presence of finite applied magnetic fields. Our results contrast with the recent observation of a seemingly two-dimensional paraconductivity in clean LiFeAs single crystals, in spite that the coherence length amplitudes of this compound are similar to the ones in optimally doped BaFe$_{2-x}$Ni$_x$As$_2$. The discrepancy may be attributed to the uncertainty in the LiFeAs normal-state contribution. 

\ack

This work was supported by the Spanish MICINN and ERDF \mbox{(No.~FIS2010-19807)}, and by the Xunta de Galicia (Nos.~2010/XA043 and 10TMT206012PR). SSS and ADA acknowledge support from CNPq and FAPERJ. The work at IOP, CAS in China is supported by NSFC Program (No. 11374011) and MOST of China (973 project: 2011CBA00110).

\end{document}